\documentstyle[prl,aps,twocolumn]{revtex}

\draft

        \def\half{{\textstyle \frac{1}{2}}}

\newcommand\binom[2]{\pmatrix{ {#1} \cr {#2}}}

\def\cP{{\cal P}}

\def\ds{\displaystyle}
 \def\ts{\textstyle}
\def\bra{\langle}
\def\ket{\rangle}

\def\rt2{\ts \frac{1}{\sqrt{2}} }

\def\uparw{\uparrow}
\def\dwnarw{\downarrow}
\def\half{{\textstyle \frac{1}{2}}}
\def\frth{{\textstyle \frac{1}{4}}}

        \title{Pauli Exchange Errors in Quantum Computation}

        \author   {Mary Beth Ruskai}
 \address{ Department of
        Mathematics, University of Massachusetts  Lowell,  Lowell,
        MA  01854 USA}

\begin{document}

\date{\today}

 \maketitle

\begin{abstract}
In many physically realistic models of quantum
        computation, Pauli exchange interactions cause a subset of
        two-qubit errors to occur as a first order effect of
        couplings within the computer, even in the absence of
        interactions with the computer's environment. We give an
        explicit 9-qubit code that corrects both Pauli exchange
        errors and all one-qubit errors.

\end{abstract}


Most schemes for fault tolerant quantum computation
        treat single qubit errors as the primary error event,
        and correct multiple-qubit errors as a higher-order side
        effect.  Discussions of quantum error
        correction also often ignore the Pauli exclusion principle
        and permutational symmetry of the states of multi-qubit
        systems. This can often be justified
        approximately by considering the full wave function,
        including spatial as well as spin components
However,  an analysis
        of these more complete wave functions suggests that exchange
        errors, in which interactions between identical particles
        cause an error in two qubits simultaneously, may be an
        important  error mechanism in some circumstances. Moreover,
        because they result from interactions within the quantum
        computer, exchange errors cannot be reduced by
        better isolating the quantum computer from its
        environment. After describing the physical mechanism of
        exchange errors, we discuss codes designed
        specifically to correct them.

A (pure) state of a quantum mechanical particle with spin $q$
corresponds to a one-dimensional subspace of the Hilbert space
${\cal H} = {\bf C}^{2q+1} \otimes L^2({\bf R}^3)$ and is typically 
represented by a vector in that subspace.   The state of a
system of $N$ such particles is then represented by a vector
$\Psi(x_1,x_2,\ldots,x_N)$ in ${\cal H}^N$.  However, when dealing
with identical particles $\Psi$ must also satisfy the Pauli
principle, i.e., it must be symmetric or anti-symmetric under
exchange of the coordinates $x_j \leftrightarrow x_k$ depending
on whether the particles in question are bosons (e.g. photons)
or fermions (e.g., electrons).  In either case, we can write
the full wave function in the form
\begin{eqnarray}
\lefteqn{ \Psi(x_1,x_2,\ldots,x_N) } \\
& = & \sum_k   \nonumber
   \chi_k(s_1,s_2,\ldots,s_N) \Phi_k({\bf r}_1,{\bf r}_2,\ldots,{\bf r}_N)
\end{eqnarray}
where the ``space functions" $\Phi_k$ are elements of 
$L^2({\bf R}^{3N})$, the ``spin functions" $\chi_k$ are in
 $[{\bf C}^{2q+1}]^N$ and $x_k = ({\bf r}_k,s_k)$
with ${\bf r}$ with a vector  in ${\bf R}^3$ and the so-called
``spin coordinat''  $s_k$  in $0, 1, \ldots 2q$. 
[In the parlance of quantum computing a spin state $\chi$  
is a (possibly entangled) N-qubit state.]   It
is not necessary that  $\chi$ and $\Phi$ each satisfy the Pauli 
principle; indeed, when $q= \half$ so that $2q+1 = 2$ and we
are dealing with ${\bf C}^2$ it is {\em not} possible for $\chi$
to be anti-symmetric when $N \geq 3$.   Instead, we 
expect that  $\chi$ and $\Phi$  satisfy certain duality conditions
which guarantee that $\Psi$ has the correct permutational symmetry.

With this background, we now restrict attention to the
important special case in which $q = \half$ yielding 
two spin states labeled so that $s = 0$ corresponds to $|0\ket$  
and  $s = 1$ corresponds to $|1\ket$, 
 and the particles are electrons so that $\Psi$ must
be anti-symmetric. 
We present our brief for the importance of Pauli exchange
errors by analyzing the two-qubit case in detail, under
the additional simplifying assumption that the Hamiltonian
is spin-free.
Analogous considerations apply in other cases.

For multi-particle states, it is sometimes convenient to
replace $|0\ket$ and $|1\ket$ by $\uparw$ and $\dwnarw$ respectively.
The notation $|01\ket$ describes a two-qubit state in which the
particle in the first qubit has spin ``up'' ($\uparrow$) and that
in the second has spin ``down'' ($\downarrow$).  What does it
mean for a particle to ``be'' in a qubit?   A reasonable answer
is that each qubit is identified by the spatial  component of its
wave function  $f_A({\bf r})$ where $A, B, C \ldots$ label the
qubits and wave functions for different qubits are orthogonal.  
Thus, 
\begin{eqnarray}\label{psi01}
  |01\ket = \rt2 \big( f_A({\bf r_1})\!\uparw \, f_B({\bf r_2})\!\dwnarw
   -   f_B({\bf r_1})\!\dwnarw \, f_A({\bf r_2})\!\uparw  \big).
\end{eqnarray} 
Notice that the electron whose spatial function is $f_A$ always
has spin ``up'' regardless of whether its coordinates are
labeled by $1$ or $2$.  We can rewrite (\ref{psi01}) as 
\begin{eqnarray}\label{eq:twobit}
|01\ket =  {\textstyle \rt2 } 
  [ \chi^+\!(s_1,s_2) \phi^-\!({\bf r_1},{\bf r_2}) +
 \chi^-\!(s_1,s_2) \phi^+ \!({\bf r_1},{\bf r_2}) ]
\end{eqnarray} 
where $\chi^{\pm}= \rt2 \left[ \uparw \dwnarw \pm \dwnarw \uparw \right]$
denote the indicated Bell states and 
$\phi^{\pm} = \rt2 \left[f_A({\bf r_1}) f_B({\bf
r_2}) \pm  f_B({\bf r_1}) f_A({\bf r_2}) \right] .$

The assumption of a spin-free  Hamiltonian $H$, implies that
the time development of  (\ref{psi01}) is determined 
 by $e^{-iHt} \phi^{\pm}$, and the assumption that the particles 
are electrons implies that $H$ includes a term corresponding
to the $\frac{1}{r_{12}} \equiv \frac{1}{|{\bf r_1}-{\bf r_2}|}$
electron-electron interaction. The Hamiltonian must be
symmetric so that the states
$\phi^{\pm}$ retain their permutational symmetry; however, the
 interaction term implies that they
 will not retain the simple form of symmetrized (or anti-symmetrized)
product states.  Hence, after some time the states $\phi^{\pm}$
 evolve into 
\begin{mathletters} \label{eq:phi.time} \begin{eqnarray}
 \Phi^- & = & \sum_{m < n} c_{mn} \rt2 \left[
  f_m({\bf r_1}) f_n({\bf r_2}) -  f_n({\bf r_1}) f_m({\bf r_2}) \right] \\
\Phi^+  & = & \sum_{m \leq n} d_{mn} \rt2 \left[
 f_m({\bf r_1}) f_n({\bf r_2}) +  f_n({\bf r_1}) f_m({\bf r_2}) \right].
\end{eqnarray} \end{mathletters} 
where $f_m$ denotes any orthonormal basis whose first two elements
are $f_A$ and $f_B$ respectively.
There is no reason to expect that $c_{mn} = d_{mn}$ in general.  
On the contrary, only the symmetric sum includes pairs with  $m = n$.
Hence if one $d_{mm} \neq 0$, then one must have some 
$c_{mn} \neq d_{mn}.$
Inserting (\ref{eq:phi.time}) in (\ref{eq:twobit}) yields 
\begin{eqnarray}
\lefteqn{ e^{-iHt} |01 \ket  = \Psi^{\rm Remain} } \nonumber \\
 & + & \frac{c_{AB} + d_{AB}}{2}
  \big( f_A({\bf r_1}) \! \uparw  f_B({\bf r_2}) \! \dwnarw 
  -  f_B({\bf r_1})\! \dwnarw   f_A({\bf r_2})\! \uparw \big) \nonumber \\
  & + &  \frac{c_{AB} - d_{AB}}{2}
  \big( f_B({\bf r_1})\! \uparw  f_A({\bf r_2})\! \dwnarw 
  -  f_A({\bf r_1})\! \dwnarw  f_B({\bf r_2})\! \uparw  \big) \nonumber \\
 & = & \frac{c_{AB} + d_{AB}}{2}|01 \ket + \frac{c_{AB} - d_{AB}}{2} |10 \ket 
   + \Psi^{\rm Remain} 
\end{eqnarray} 
where $\Psi^{\rm Remain}$ is orthogonal to $\phi^{\pm}$.

A measurement of qubit-A  corresponds to projecting onto  $f_A$.  
Hence a measurement of qubit-A on the state (\ref{eq:twobit})
yields spin ``up'' with probability $\frth |c_{AB} + d_{AB}|^2$
and spin down with probability  $\frth|c_{AB} - d_{AB}|^2$, 
and zero with probability $\| \Psi^{\rm Remain} \|^2$.   Note
 that the {\em full} wave function is {\em necessarily} an {\em entangled}
state and that the measurement process leaves the system in
state $|10 \ket$ or $|01 \ket$ with probabilities 
 $\frth|c_{AB} \pm d_{AB}|^2$ respectively, i.e., subsequent measurement
of qubit-B always gives the opposite spin.  With probability
 $\frth|c_{AB} - d_{AB}|^2$ the initial state $|10 \ket$ has been
converted to $|01 \ket$. 

Although the probability of this may be small, it is {\em not} zero.  
Precise estimates require a more detailed model
of the actual experimental implementation.  
However, it would seem that any
implementation which provides a mechanism for two-qubit
gates would necessarily permit some type of
interaction between particles in different qubits.
Because one expects qubits to be less isolated from
each other than from the external environment,
 Pauli exchange errors seem  to merit more attention.

If the implementation involves charged particles,
whether electrons or nuclei, then the interaction includes a
contribution from the $\frac{1}{r_{12}}$ Coulomb potential
which is known to have long-range effects.  This suggests
that implementations involving neutral particles, such as
Briegel, et al's  proposal \cite{BCJCZ}  using optical lattices, may be
advantageous for minimizing exchange errors.

A Pauli exchange error is a special type of ``two-qubit" error
 which has the same effect as ``bit flips" if (and {\it only} if)
they are different.  Exchange of bits  $j$ and $k$ is equivalent
to acting on a state with the operator
\begin{eqnarray}\label{def:Ejk}
  E_{jk} = \half \Big( I_j \otimes I_k + Z_j \otimes Z_k
   + X_j \otimes X_k  + Y_j \otimes Y_k \Big)
\end{eqnarray}
where $X_j, Y_j, Z_j$ denote the action of the Pauli matrices
$\sigma_x,  \sigma_y, \sigma_z$ respectively on the bit $j$.

%pagebreak  
As an example, we consider  Pauli exchange errors
in the simple 9-bit code of  Shor \cite{Shor}
   \begin{mathletters} \label{Shorcode} \begin{eqnarray} 
 %    \begin{eqnarray} \label{Shorcode}
| c_0 \ket    & = & |{\bf 000} \ket  + |{\bf 011} \ket + 
   |{\bf 101} \ket  + |{\bf 110} \ket  \\
 | c_1 \ket & = &    |{\bf 111} \ket  +
    |{\bf 100} \ket + |{\bf 010}\ket  + |{\bf 001} \ket 
\end{eqnarray}\end{mathletters}
%\end{eqnarray}
where boldface denotes a triplet of 0's or 1's.
It is clear that these code words are invariant under exchange
of electrons within the 3-qubit triples (1,2,3),  (4,5,6), or (7,8,9).
To see what happens when electrons in different triplets are
exchanged, consider the exchange $E_{34}$ acting on $| c_0 \ket $.
This yields $ |000000000\ket + |001011111\ket +
|110100111\ket  +  |111111000\ket $ so that
\begin{eqnarray*}
  E_{34}  | c_0  \ket & = &  | c_0 \ket +  Z_2  | c_0 \ket +
    |001011111\ket + |110100111\ket   \\
      E_{34} | c_1 \ket  & = &  | c_1 \ket -  Z_2  | c_1 \ket 
    + |110100000\ket + |001011000 \ket  
\end{eqnarray*} 
If $|\psi \ket = a | c_0 \ket + b | c_1 \ket$ is a superposition
of code words, 
\begin{eqnarray*} 
   E_{34} |\psi \ket  = \half \Big( |\psi \ket +  
     Z_8 | \tilde{\psi} \ket  \Big) + \rt2 | \gamma \ket
\end{eqnarray*}
where $ | \tilde{\psi} \ket = a | c_0 \ket - b | c_1 \ket$ differs
from $\psi$ by a ``phase error" on the code words and  $| \gamma \ket$
is orthogonal to the space of codewords and single bit errors.
Thus, this code cannot reliably distinguish between an exchange
error $E_{34}$ and a phase error on any of the last 3 bits.  This
problem occurs because if
$ E_{34} | c_0 \ket = \alpha  | c_0 \ket + \beta | d_0 \ket $
with $| d_0 \ket $  orthogonal to $| c_0 \ket$, then  $| d_0 \ket $ need
not be orthogonal to $| c_1 \ket$.

In order to be able to correct a given class of errors, we first identify
a set of basic errors $e_p$  in terms of which all other errors can
be written as linear combinations.  In the case of unitary transformations
on single bit, or one-qubit errors, this set usually consists of
$X_k, Y_k, Z_k ~~ (k= 1 \ldots n)$ where $n$ is the number of qubits in
the code and $X_k, Y_k, Z_k$ now denote 
$I \otimes I \otimes I \ldots\otimes  \sigma_p \otimes \ldots \otimes I$
where $ \sigma_p $ denotes one of the three Pauli matrices acting
on qubit-k.  If we let $e_0 = I$ denote the identity, then a
sufficient condition for error correction is
\begin{eqnarray}\label{err.orthog}
\bra e_p C_i | e_q C_j \ket = \delta_{ij} \delta_{pq}
\end{eqnarray}
However, (\ref{err.orthog})  can be replaced 
\cite{CRSS,KL} by the weaker 
\begin{eqnarray}\label{err.suff}
  \bra e_p C_i | e_q C_j \ket = \delta_{ij} d_{pq} .
\end{eqnarray}
where the matrix $D$ with elements $d_{pq}$ is independent of $i,j$.
When considering Pauli exchange errors, it is natural to seek codes
which are invariant under some subset of permutations.  This is
clearly incompatible with (\ref{err.orthog}) since some of the exchange
errors will then satisfy $ E_{jk} | C_i \ket = | C_i \ket $.
Hence we will need to use (\ref{err.suff}).

The most common code words have the
property that $| C_1 \ket $ can be obtained from $| C_0 \ket$
by exchanging all 0's and 1's. For such codes, it is not
hard to see that
 $ \bra C_1 | Z_k C_1 \ket =  - \bra C_0 | Z_k C_0 \ket $
which is consistent with (\ref{err.suff}) if and only if 
it is identically zero.
Hence even when using (\ref{err.suff}) rather than (\ref{err.orthog})
it is necessary to require 
\begin{eqnarray}\label{eq:dualphase}
  \bra C_1 | Z_k C_1 \ket =  - \bra C_0 | Z_k C_0 \ket = 0
\end{eqnarray}
when the code words are related in this way.

We now  present  a 9-bit code code which  
can handle both  Pauli exchange errors and all one-bit errors.
It is based on the realization that codes
which are invariant under permutations are impervious to
Pauli exchange errors. 
Let 
\begin{mathletters} \label{code9} \begin{eqnarray}
% \begin{eqnarray} \label{code9} 
| C_0 \ket & = &  |000\, 000\, 000\ket + 
    \frac{1}{\sqrt{28}} \sum_{\cP} |111\,111\,000\ket \\
 | C_1 \ket & = &  |111\,111\,111\ket + 
     \frac{1}{\sqrt{28}} \sum_{\cP} |000\,000\,111\ket 
\end{eqnarray} \end{mathletters}
% \end{eqnarray} 
where $\sum_{\cP}$ denotes the sum over all permutations of the
indicated sequence of 0's and 1's and it is understood that we
count permutations which result in identical vectors only once.
This differs from the  9-bit Shor code  in that 
{\em all} permutations of $|111\,111\,000\ket$ are included,
rather than only three.  The normalization of the code words is
$\bra  C_i |  C_i \ket  = 1 + \frac{1}{28} \pmatrix{9 \cr 3} = 4$.

The coefficient $1/\sqrt{28}$ is needed to satisfy (\ref{eq:dualphase}).
Simple combinatorics implies
\begin{eqnarray*}
\bra  C_i | Z_k C_i \ket = 
  (-1)^i \left[ 1 - \frac{1}{3}\binom{9}{3}\frac{1}{28} \right]
  = 0.
\end{eqnarray*}
Moreover,
\begin{eqnarray}\label{eq:phase.mat}
\bra  Z_k C_i | Z_{\ell} C_i \ket = 
  1 + \delta_{k \ell} \binom {9}{3}\frac{1}{28}  = 1 + 3 \delta_{k \ell} .
\end{eqnarray}
The  second term in (\ref{eq:phase.mat}) is zero 
when $k \neq \ell$ because of
the fortuitous  fact that there are exactly the same
number of positive and negative terms.  If, instead, we had
used all permutations of $\kappa$ 1's in $n$ qubits, this term would be
 $\ds{\frac{(n-2 \kappa)^2 - n}{n(n-1)} }\binom{n}{\kappa}$ when
$k \neq \ell$.
 
Since all components of $|C_0\ket$ have $0$ or  $6$ bits equal to 1, 
any single bit flip acting on $|C_0\ket$, will yield a vector
whose components have $1,  5$, or $7$ bits equal to 1 and is
thus orthogonal to $|C_0\ket$, to $|C_1\ket$, to a bit flip acting
on $|C_1\ket$ and to a
phase error on either  $|C_0\ket$ or $|C_1\ket$.
Similarly, a single bit flip on $|C_1\ket$ will yield a vector
orthogonal to $|C_0\ket$, to $|C_1\ket$, to a bit flip acting
on $|C_0\ket$ and to a  phase error on $|C_0\ket$ or $|C_1\ket$. 
 However, single bit flips on a given code word
are not mutually orthogonal.

To find $\bra  X_k C_i | X_{\ell} C_i \ket$ when $k \neq \ell$,
consider
\begin{eqnarray}\label{eq:bit.mat}
\bra  X_k \, (\nu_1 \nu_2 \ldots \nu_9) \, | \,
                 X_{\ell} \, (\mu_1 \mu_2 \ldots \mu_9) \ket. 
\end{eqnarray}
where $\nu_i, \mu_i$ are in $0,1$. 
This will be nonzero only when 
$\nu_k = \mu_{\ell} = 0, ~~  \nu_{\ell} =  \mu_k = 1$ or
$\nu_k = \mu_{\ell} = 1, ~~  \nu_{\ell} =  \mu_k = 0$
and the other $n-2$ bits are equal.  From $\sum_{\cP}$ with 
$\kappa$ of $n$ bits
equal to 1, there are $2 \binom{n-2}{\kappa-1}$ such terms.  Thus, for the
code (\ref{code9}), there are 42 such terms which yields an
inner product of $\frac{42}{28} = \frac{3}{2}$ when $k \neq \ell$.
If we consider instead, 
$\bra  Y_k C_i | X_{\ell} C_i \ket 
  = - i \bra  X_k Z_k C_i | X_{\ell} C_i \ket$
for $k \neq \ell$ it is not hard to see that exactly half of the
terms analogous to (\ref{eq:bit.mat}) will occur with a positive
sign and half with a negative sign, yielding  a net  inner product
of zero.
We also find 
$\bra  Y_k C_i | X_k C_i \ket = - i \bra  X_k Z_k C_i | X_k C_i \ket
  = -i \bra  Z_k C_i |  C_i \ket = 0 $
so that $\bra  Y_k C_i | X_{\ell} C_i \ket = 0 $ for all $k, \ell$.
In addition
$\bra  Y_k C_i | Z_{\ell} C_i \ket = 
 - i \bra  X_k Z_k C_i | Z_{\ell} C_i \ket
   = 0 $
for the same reason that $\bra  X_k C_i |  C_i \ket = 0$.

These  results imply that  (\ref{err.suff}) holds and 
that the matrix  $D$ is block diagonal with the form
\begin{eqnarray} 
%D = \left( \begin{array}{cccc}  
D = \pmatrix{   D_0 & 0  & 0 & 0 \cr
   0 &  D_X & 0 & 0 \cr
   0 & 0 & D_Y & 0  \cr
   0 & 0 & 0 & D_Z }
%\end{array} \right)
\end{eqnarray} 
where $D_0$ is the $37 \times 37$ matrix corresponding to the 
identity and the 36 exchange errors, and
$D_X, D_Y, D_Z$ are $9 \times 9$ matrices corresponding
respectively to the $X_k, Y_k, Z_k$ single bit errors.
One easily finds that $d^0_{pq} = 4$ for all $p,q$.
The $9 \times 9$ matrices $D_X, D_Y, D_Z$ all have 
$d_{kk} = 4$ while for $ k \neq \ell$, $ d_{k \ell} =  3/2$ 
in $D_X$ and $D_Y$ but $d_{k \ell}  = 1$ in $D_Z$.
Orthogonalization of this matrix is straightforward.
Since $D$ has rank $28 = 3 \cdot 9 + 1$, we are using only
a $54 < 2^6$ dimensional subspace of our $2^{9}$ dimension space.

The simplicity of codes which 
 are invariant under permutations makes them attractive.
However, there are few such codes.   All code words necessarily have the form
\begin{eqnarray}\label{eq:permcode}
  \sum_{\kappa=0}^n a_\kappa \sum_{\cP} 
   |\underbrace{ 1  \ldots 1}_\kappa \underbrace{ 0 \ldots 0}_{n-\kappa} \ket.
\end{eqnarray}
 Condition 
(\ref{err.suff})  places some severe restrictions
on the coefficient $a_\kappa.$  For example, in (\ref{code9}) 
only $a_0$ and $a_6$
are non-zero in $|C_0\ket$ and only $a_3$ and $a_9$ in $|C_1\ket$.
If we try to change this so that $a_0$ and $a_3$ are non-zero
in $|C_0\ket$ and  $a_6$ and $a_9$ in $|C_1\ket$, then it is {\em not}
possible to satisfy (\ref{eq:dualphase}).

The 5-bit code in \cite{CRSS,LMPZ,BDSW} does not have the 
permutationally invariant form
(\ref{eq:permcode}) because the code words include components
of the form $\sum_{\cP} \pm | 11000 \ket $, i.e., not all terms
in the sum have the same sign.
 The non-additive 5-bit code in  \cite{RHSS}  requires
sign changes in the $\sum_{\cP} \pm | 10000 \ket$ term.
Since such sign changes seem needed to satisfy (\ref{eq:dualphase}),
it appears that that 5-bit codes can  not handle 
Pauli exchange errors (although we have no proof).

However, permutational invariance, which is based on
a one-dimensional representation of the symmetric group, 
is not the only approach to exchange errors.  Our analysis of 
(\ref{Shorcode}) suggests a construction which we
first describe in over-simplified form.   Let 
$|c_0 \ket, |d_0 \ket, |c_1 \ket, |d_1 \ket$  be four  mutually orthogonal
n-bit vectors such that $|c_0 \ket,  |c_1 \ket$ form a  code
for one-bit errors and $|c_0 \ket, |d_0 \ket$ and $|c_1 \ket, |d_1 \ket$
are each bases of a two-dimensional representation of the symmetric
group $S_n$.  If $|d_0 \ket$ and $|d_1 \ket$  are also
orthogonal to one-bit errors on the code words, then this code can correct
Pauli exchange errors as well as one-bit errors.  If, in addition,
the vectors $|d_0 \ket,  |d_1 \ket$ also form a code
isomorphic to $|c_0 \ket,  |c_1 \ket$ in the sense that the matrix
$D$ in (\ref{err.suff}) is identical for both codes, then the code
should also be able to correct products of one-bit and 
Pauli exchange errors.

But the smallest (excluding one-dimensional) irreducible representations 
of the  symmetric group 
for use with n-bit codes have dimension $n-1$.
Thus we will seek a set of $2(n-1)$ mutually orthogonal vectors denoted
$|C_0^m \ket,  |C_1^m \ket ~(m= 1 \ldots n-1)$ such that
$|C_0^1 \ket,  |C_1^1 \ket$ form a code for one bit errors and
$|C_0^m \ket ~(m= 1 \ldots n-1)$ and $|C_1^m \ket ~(m= 1 \ldots n-1)$ 
each form basis of the same irreducible representation of $S_n$.
Such code will be able to correct {\em all} errors which permute
qubits; not just single exchanges.
If, in addition, (\ref{err.suff}) is extended to 
\begin{eqnarray}\label{err.suff.ext}
  \bra e_p C_i^m | e_q C_j^{m'} \ket = \delta_{ij} \delta_{mm'} d_{pq}
\end{eqnarray}
with the matrix $D = \{D_{pq}\}$ independent of both $i$ and $m$,
then this code will also  be able to correct 
 products of one bit errors and permutation errors.

If  the basic error set has size
$N$ (i.e., $p = 0, 1 \ldots N-1$), then a two-word code requires
codes which lie in a space of dimension at least $2N$.  For 
the familiar case of single-bit errors $N = 3n+1$ and, since an
n-bit code word lies in a space of dimension $2^n$, any code must
satisfy $3n+1 < 2^{n-1}$ or $n \geq 5$.
There are $n(n-1)/2$ possible single exchange
errors compared to $9n(n-1)/2$ two-bit errors of all types.  
Similar dimension arguments yield
 $2N = n^2 + 5n +2 \leq 2^n$ or $n \geq 7$ for correcting both
single-bit and single-exchange errors and
 $2N = 9n(n-1) + 2(3n+1) \leq 2^n$ or $n \geq 10$ for correcting
all one- and two-bit errors.   The shortest code  known \cite{CRSS}
which can do the latter has n = 11.  
Correcting Pauli exchange errors can be done with shorter
codes than required to correct all two-bit errors.  

However, this simple dimensional analysis need not yield the best
bounds when exchange errors are involved.  Consider
 the simple code $ | C_0 \ket =  |000\ket, | C_1 \ket =  |111\ket $
which is optimal for single bit flips (but
can not correct phase errors).  In this case $N = n+1$ and $n = 3$
yields equality in  $2(n + 1) \leq 2^n$.
But, since this code is invariant
under permutations, the basic error set can be
expanded to include all 6 exchange errors $E_{jk}$ for a total
of $N = 10$ without increasing the length of the code words.

In the construction proposed above, correction of exchange and one-bit errors
would require a space of dimension $2(n-1)(3n+1) \leq 2^n$ or
$n \geq 9$.  If codes satisfying (\ref{err.suff.ext}) exist, 
 they could correct {\em all} permutation errors
as well as products of permutations and one-bit errors.   
Exploiting permutational symmetry  may have a big payoff.

Although codes which can correct Pauli exchange errors will be
larger than the minimal 5-qubit codes proposed for single-bit
error correction, this may not be a serious drawback.  
 For implementations of quantum computers which have a grid
structure (e.g., solid state or optical lattices)  it may be natural and
advantageous to use 9-qubit codes  which can be implemented in
$3 \times 3$ blocks. (See, e.g.,  \cite{BCJCZ}.)   
However, codes larger than 9-bits may be impractical
for a variety of reasons.    Hence it is encouraging that both the 
code (\ref{code9})  and the proposed construction above
do not require $n > 9$.

Several more complex coding schemes have been proposed 
\cite{KL,KLP,Kt,St,KLV,DFS}
for multiple error correction.  It may be worth
investigating whether or not the codes proposed here 
can be used advantageously in some  of these schemes, 
such as those \cite{KLP} based on  hierarchical nesting.  
Since the code (\ref{code9}) can already handle  multiple exchange errors, 
(and the proposed construction some additional multiple errors)
concatenation of one of our proposed 9-bit codes with itself
will contain some redundancy and concatenation with a 5-bit code
may be worth exploring.

Whether or not any 7-bit codes exist which can handle Pauli exchange
errors is another open question, which we leave as another challenge
for coding theorists.

\bigskip

\noindent{\bf Acknowledgment}  It is a pleasure to thank Professor
Eric Carlen for a useful comment and both Professor Chris King 
and Professor Harriet Pollatsek for
several helpful discussions and comments on earlier drafts.

This research was partially supported by National Science
        Foundation Grant DMS-97-06981 and Army Research Office Grant
   DAAG55-98-1-0374.

\end{document}